\documentclass[twocolumn,journal]{IEEEtran}
\usepackage{epsf}
\usepackage{graphicx}
\usepackage{epsfig,latexsym,amsmath,epsf,amssymb,amsfonts}
\usepackage{amssymb}
\usepackage{placeins}
\usepackage{epsfig}
\usepackage{epstopdf}
\usepackage{color}
\usepackage{lipsum}
\usepackage{balance}
\usepackage{graphicx}
\usepackage{amsmath}
\usepackage{amssymb}
\usepackage{balance}

\usepackage{amsthm}
\usepackage{cite}
\usepackage{float}
\usepackage{color}

\begin{document}

\title{Interference-Based Optimal Power-Efficient Access Scheme for Cognitive Radio Networks}

\author{\authorblockN{\small{Ahmed H. Anwar\authorrefmark{1}, Ahmed El Shafie\authorrefmark{2}, Amr Mohamed\authorrefmark{1}, Tamer ElBatt\authorrefmark{2}} and Mohsen Guizani\authorrefmark{1}}\\
\authorblockA{\small{\authorrefmark{1}Computer Science and Engineering Dept., Qatar University, Doha, Qatar.}\\ \authorblockA{\small{\authorrefmark{2}Wireless Intelligent Networks Center (WINC), Nile University, Giza, Egypt.}} \\  \small{
email: \{a.h.anwar@knights.ucf.edu, ahmed.salahelshafie@gmail.com, amrm, telbatt@ieee.org and mguizani@qu.edu.qa\} }}}

\maketitle

\begin{abstract}
In this paper, we propose a new optimization-based access strategy of multipacket reception (MPR) channel for multiple secondary users (SUs) accessing the primary user (PU) spectrum opportunistically. We devise an analytical model that realizes the multipacket access strategy of SUs that maximizes the throughput of individual backlogged SUs subject to queue stability of the PU. All the network receiving nodes have MPR capability. We aim at maximizing the throughput of the individual SUs such that the PU's queue is maintained stable. Moreover, we are interested in providing an energy-efficient cognitive scheme. Therefore, we include energy constraints on the PU and SU average transmitted energy to the optimization problem. Each SU accesses the medium with certain probability that depends on the PU's activity, i.e., active or inactive. The numerical results show the advantage in terms of SU throughput of the proposed scheme over the conventional access scheme, where the SUs access the channel randomly with fixed power when the PU is sensed to be idle.
\end{abstract}
\footnotetext[1]{This work was made possible by grant number NPRP 5-782-2-322 and NPRP 4-1034-2-385 from the Qatar National Research Fund (a member of Qatar Foundation). The statements made herein are solely the responsibility of the authors.}
\section{Introduction}
\IEEEPARstart{C}\small{ognitive} radio networks (CRN) have been a hot area of research for a decade due
to its opportunistic, agile and efficient spectrum utilization
merits \cite{mitola1999cognitive,haykin2005cognitive,akyildiz2006next}. Cognitive radios, in the literature, promote three methods of sharing the spectrum with the primary user (PU); namely, overlay, underlay and interweave schemes \cite{goldsmith2009breaking}. Under collision wireless channel model, the authors of \cite{luo1999stability,rao1988stability} studied the stability of multiple access slotted ALOHA systems. The authors used the concept of dominant systems to characterize the stability region and provide sufficient and necessary conditions for stability of the system.  Time is slotted and each node randomly accesses the channel. In \cite{luo2006throughput}, the authors investigated the stability region, capacity and throughput for multiple access ALOHA systems.

In \cite{sadek}, Sadek {\it et al.} investigated the stability region of a network composed of a cognitive relay that aids multiple nodes for the transmission of their packets to a common destination. The proposed cooperative protocols enable the relaying node to aid the transmitters operating in a time-division multiple access (TDMA) network in their silent periods due to
source burstiness. In \cite{el2011opportunistic}, El-Sherif investigated the stability region of a network composed of multiple relays cooperate in forwarding the traffic of multiple PUs during their silence periods with the coexistence of multiple secondary nodes. The authors considered two secondary access scenarios which give inner and outer bounds to the original system's performance. Under the first scenario, the secondary users (SUs) discern the activity of both the PUs and the relays at each time slot, thereby remaining silent when any of them is active. Under the second scenario, the relays and the SUs randomly access the channel every time slot and thus transmissions may collide. Consequently, packet loss may occur. In \cite{TVT}, the authors propose two order cognitive access schemes which differ in terms of the required coordination between the secondary terminals. Under the proposed schemes, the secondary terminals are ordered in terms of channel accessing. The proposed protocol is studied from the network layer point of view with a collision-based wireless channel model.

In \cite{ghez1988stability}, the authors introduced a generalized channel model with multipacket reception (MPR) capability for slotted ALOHA systems where receivers are capable of decoding under interference. They investigated the stability of the system compared to the collision channel model. Authors in \cite{naware2005stability} studied the impact of MPR on stability
and delay of slotted ALOHA-based random access systems. The stability region is characterized using dominant systems approach. Kompella {\it et al.} in \cite{erph}, characterized the stable-throughput region of an SU sharing the channel with a PU. The PU has unconditional access to the channel, while the SU transmits its packets with some access probability that changes based on the primary queue state. Precisely, if the PU is inactive, the SU accesses the channel unconditionally, and if the PU is active, the SU accesses with some probability. The channel sensing is assumed to be perfect. In \cite{wimob}, the authors investigated a cognitive setting with one PU and one rechargeable SU. The SU randomly accesses and senses the primary channel and can possibly leverage the primary feedback. Receivers are capable of decoding under interference as they have MPR capabilities. The authors investigated the maximum secondary throughput under the PU stability and delay constraints. In \cite{ourletter}, the SU randomly accesses the channel at the beginning of the time slot to exploit the MPR capability of receivers. The SU aims at maximizing its throughput under PU's queue stability and certain queueing delay requirement for the PU.

Energy efficient protocols are of a great importance currently due to the huge demand on applications and communications of limited energy. In this paper, we focus on energy-efficient and power-limited communication systems; hence, we consider an underlay cognitive scheme that matches our interests in studying both the PU and SUs energy and interference effects as explained later. Using cognitive networks in the field of ``{\it Green Communications}'' was introduced earlier by \cite{grace2009using}. A recent survey on energy-efficient wireless communications and new protocols are presented in \cite{feng2012survey}. The authors of \cite{lim2012energy} discussed the energy-efficient relay selection techniques in a cooperative heterogeneous radio access network. Fundamental trade-offs and challenging problems regarding green communications are explained in \cite{hossain2012green}.

In this paper, we devise a new model that captures energy efficiency and throughput optimization. Our contributions can be summarized as follows:
\begin{itemize}
\item We propose a new access scheme that maximizes throughput of individual SUs, while guarantees the stability of the PU's queue and limits the transmission powers of SUs and PU.
\item We show that the SUs throughput increase via controlling the optimization problem variables.
\item We conduct a comprehensive study to show the effect of the system parameters on both throughput maximization and power efficiency from a design perspective.
\end{itemize}

The rest of the paper is organized as follows. The system model is presented in Section \ref{System}. Then the throughput is characterized in Section \ref{Throughput}. We perform the optimization problem in Section \ref{Optimization}. Numerical results are presented in Section \ref{Results}. Finally, we conclude our work in Section \ref{Concl}.

\section{System Model}\label{System}

\begin{figure}
\centering
\includegraphics[width=0.9\columnwidth]{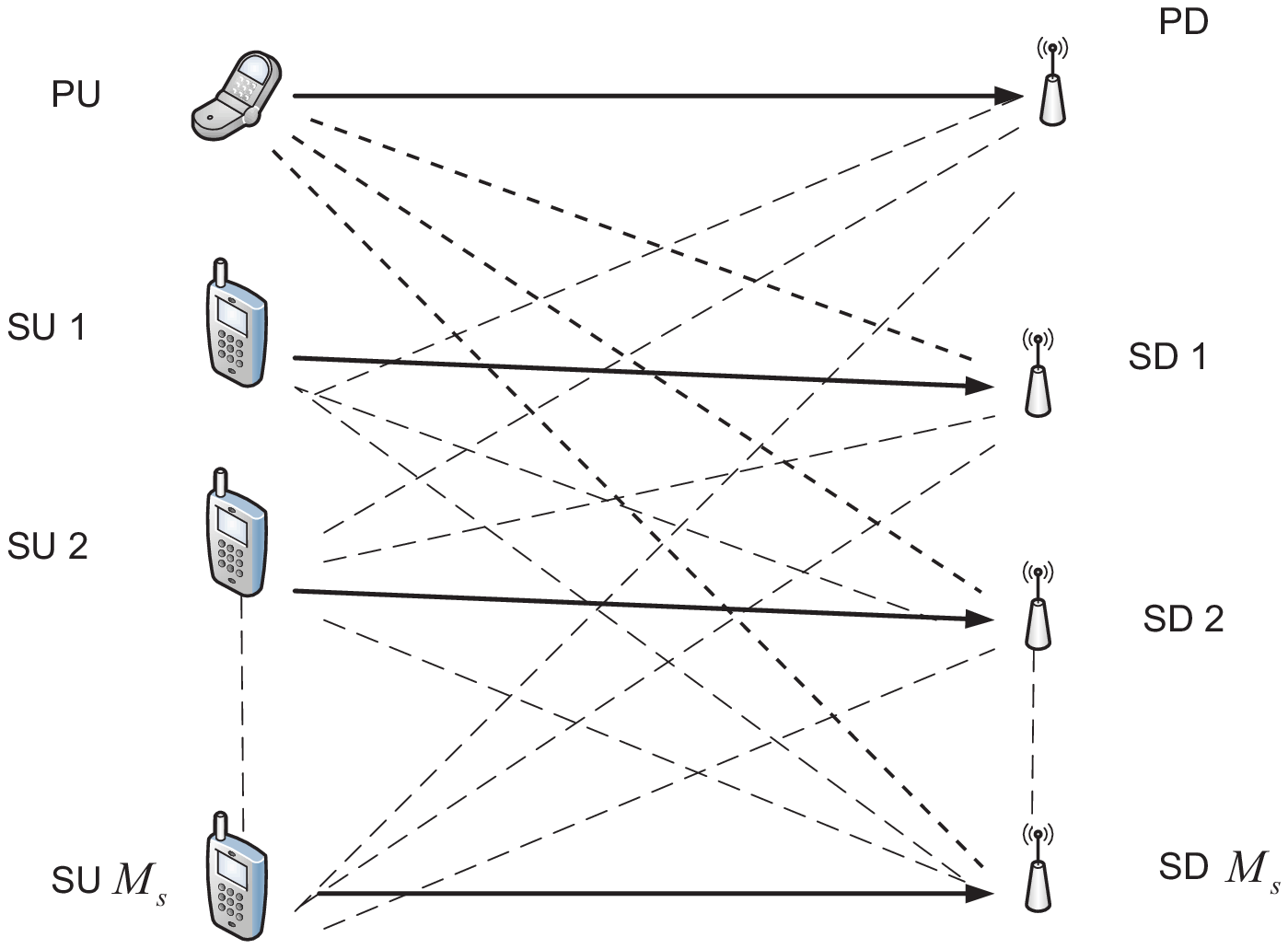}
\caption{Primary and secondary links. The solid links represent the communication links, while the dotted links represent the interference links. In the figure, we denote the primary and secondary destinations by PD and SD, respectively.}\label{model}
\end{figure}

In this paper, we consider a single primary link and $M_{\rm s}$ secondary links sharing one frequency channel as shown in Fig. \ref{model}. Time and channels are slotted. Each time slot is $T$ seconds in length. The PU has a total transmission bandwidth of $W$ Hz. The primary transmitter has a buffer (queue) $Q_{\rm p}$ modeled as Geo/Geo/1 queueing system. The arrivals at the primary queue are assumed to be independent and identically
distributed (i.i.d) Bernoulli random variables with stationary mean $\lambda_{\rm p}\in[0,1]$ packets per time slot. We adopt a late-arrival model for the primary queue, where a packet cannot be served in the arrival time slot even if the queue is empty. Each secondary transmitter has a buffer for storing its arrived packets, denoted by $Q_{j}$, $j\in\{1,2,\dots,M_{\rm s}\}$. The SUs are assumed to be saturated (always backlogged) with data packets, i.e., each SU has a data packet to send every time slot. All buffers of the system have infinite capacity \cite{sadek,erph,ourletter}.

We assume that the SUs sense the medium perfectly each time slot to detect the activity of the PU as in \cite{erph} and the references therein. This assumption is practically valid if the channel sensing duration is long enough to guarantee negligible sensing errors which is the case we have. All SUs randomly access the medium every time slot with certain access probability that changes based on the state of the sensed PU. Typically, if the medium is sensed to be idle, the $j^{\it th}$ SU accesses the medium with probability $a^{\left(j\right)}_1$ and power $\gamma^{\left(j\right)}_1$. On the other hand, if the medium is sensed to be busy, the $j^{\it th}$ SU accesses the medium with probability $ a^{\left(j\right)}_2 $ and power $ \gamma^{\left(j\right)}_2 $ such that, $ a^{\left(j\right)}_1 \geq a^{\left(j\right)}_2 $. The basic idea of our proposed scheme relies on the fact that if all users or a group of them simultaneously transmit a packet, the receivers still can decode the packets with a non-zero probability. This occurs due to the availability of the MPR capability at the receiving nodes.

For the simplicity of our presentation, we assume that all SUs are symmetric (a similar assumption is found in the literature, e.g., \cite{el2011opportunistic}). This implies that all channels have similar distribution for channels and all SUs transmit with the same power levels and access the channel with equal access probabilities. Specifically, the statistics of the channel gains for all SU are equal. Moreover, $\gamma^{\left(j\right)}_\ell=\gamma^{\left(i\right)}_\ell$ and $a^{\left(j\right)}_\ell=a^{\left(i\right)}_\ell$ for all $j,i\in\{1,2,\dots,M_{\rm s}\}$, $j\ne i$ and $\ell \in \{1,2\}$.

The SUs are energy-aware nodes that aim at efficiently expending their energy to maximize their performance and to satisfy certain quality of service (QoS) requirements for the PU. The SUs sense the medium for $ \tau $ seconds from the beginning of each time slot to determine the state of the PU. If a terminal transmits during a time slot, it sends exactly one data packet whose length is $b$ bits. Accordingly, the transmission rate of the secondary terminals is $ \tilde r_{\rm s}=b/(T-\tau)$. The PU accesses the channel unconditionally at the beginning of the time slot if its queue is non-empty; hence, the primary transmission rate is $r_{\rm p}=b/T$. The transmit power of the PU is $\gamma_{\rm p}$ Watts/Hz.

We assume block Rayleigh fading channels, where the channel gain from node $n_1$ to node $n_2$, denoted by $h_{n_1,n_2}$, is assumed to be fixed during one slot, but changes from time slot to another according to exponential distribution with mean $1/\delta_{n_1,n_2}$. Let ${\rm pd}$ and ${\rm sd}_j$ denote the primary destination and the $j^{th}$ SU's destination, respectively. The set of transmitters is given by $n_1\in\{{\rm p},1,2,\dots,M_{\rm s}\}$, whereas the set of receivers is given by $n_2\in\{{\rm pd,sd_1,sd_2,sd_3},\dots,{\rm sd}_{M_{\rm s}}\}$. Due to symmetry, $\delta_{{\rm s}_j,{\rm sd}_j}=\delta_{\rm ss}$, $\delta_{{\rm p},{\rm sd}_j}=\delta_{\rm ps}$, $\delta_{{\rm s}_j,{\rm pd}}=\delta_{\rm sp}$, and $\delta_{{\rm p},{\rm pd}}=\delta_{\rm pp}$. The thermal noise at any receiver is modeled as additive white Gaussian noise (AWGN) with zero mean and power spectral density $\mathcal{N}_\circ$ Watts/Hz.

The medium access scheme can be summarized as follows:
\begin{itemize}
\item The PU accesses the channel at the beginning of the time slot if it has a packet at the head of its queue.
\item The SUs sense the channel over the time interval $[0,\tau]$ to declare the state of the PU.
\item If the PU is idle, each SU randomly accesses the channel with probability $a_1$ and transmit power level $\gamma_1$ Watts/Hz.
\item If the PU is active, each SU randomly accesses the channel with probability $a_2$ and transmit power level $\gamma_2$ Watts/Hz.
\item At the end of the time slot, each receiving node sends back a feedback signal to the respective transmitter to declare the state of the decodability of the transmitted packet. If the packet is decoded correctly, the respective receiver sends back an acknowledgement (ACK). If the packet is decoded erroneously, due to channel outages, the respective receiver sends back a negative-acknowledgement (NACK), and the packet will be retransmitted in the following time slots.
\end{itemize}

\section{Throughput Analysis and Problem formulations}\label{Throughput}

Conventionally queue stability is considered as a fundamental performance metric in any communication system. Specifically, suppose $Q^{\left(t\right)}$ denotes the length of queue $Q$ at the beginning of time slot $t$. $Q$ is said to be stable if $\lim_{\mathcal{I} \rightarrow \infty  } \lim_{t \rightarrow \infty  } {\rm Pr}(Q^{\left(t\right)}<\mathcal{I})=1$~\cite{sadek}, where $\mathcal{I}$ is a positive constant and $ {\rm Pr}(\cdot)$ represents the probability of the argument event. If the queues are characterized with strictly stationary arrival and departure processes, we can apply Loynes criterion to check the stability of each queue \cite{sadek}. This theorem states that if the arrival process and the service process of a queue are strictly stationary, and the average service rate is greater than the average arrival rate of the queue, then the queue is stable.

Denote by $\mathcal{X}^t_{\rm p}$ the number of arrivals to queue $Q_{\rm p}$ in time slot $t$, and $\mathcal{S}^t_{\rm p}$ the number of departures from queue $Q_{\rm p}$ in time slot $t$. The queue length evolves according to the following form:
\begin{equation}
    Q_{\rm p}^{t+1}=(Q_{\rm p}^t-\mathcal{S}^t_{\rm p})^+ +\mathcal{X}^t_{\rm p}
\end{equation}
where $(z)^+=\max(z,0)$ denotes the maximum between $z$ and $0$. The queue size is measured at the beginning of the time slot, and departures occur before arrivals \cite{sadek}.

Let $ \mu _{\rm s}=\mu^j _{\rm s}$ denote the mean service rate of an SU and $ \mu_{\rm p} $ denote the mean service rate of the PU.
The mean service rate of the $j^{th}$ SU is given by

\begin{equation}\label{eqn_mu_s}
\begin{split}
& \mu_{\rm s} \!=\! \Pr \left ( Q_{\rm p}\!=\!0  \right ) \left [ \sum_{k=0}^{M_{\rm s}-1} {M_{\rm s} \choose k} a_1^k(1\!-\!a_1)^{M_{\rm s}-k} P^{\rm s}({\rm Success}|k) \!\right ] \\& \!+\!  \Pr \left (\! Q_{\rm p}\!\neq\! 0  \right ) \!\left [\! \sum_{k=0}^{M_{\rm s}-1}  {M_{\rm s} \choose k} a_2^k(\!1\!-\!a_2\!)^{M_{\rm s}\!-\!k} P^{\rm s}({\rm Success}|k\!+\!{\rm PU}) \!\right ]
\end{split}
\end{equation}

\noindent
where $k\!+\!1$ is the number of active SUs in a certain time slot, i.e., $k$ active SUs plus the $j^{\it th}$ user, $ P^{\rm s}({\rm {\rm Success}}|k) $ is the probability of decoding the $j^{\it th}$ SU packet successfully given that $ k $ out of $ M_{\rm s}-1$ SUs are accessing the medium simultaneously with the $j^{\it th}$ SU while the PU is idle, and $ P^{\rm s}({\rm {\rm Success}}|k+{\rm PU}) $ is the probability of successful packet decoding of the $j^{\it th}$ SU when $k$ SUs and the PU are accessing the channel at the same time.

Next, we characterize the different probabilities in equation (\ref{eqn_mu_s}). The probability of the primary queue being empty is given by
\begin{equation}\label{eqn_pr_Q}
\Pr(Q_{\rm p}=0) = 1-\Pr(Q_{\rm p}\neq 0 ) = 1 - \frac{\lambda_{\rm p}}{\mu_{\rm p}},
\end{equation}
\noindent
where
\begin{equation}\label{eqn_mu_p}
\mu_{\rm p}= \sum_{k=0}^{M_{\rm s}} {M_{\rm s} \choose k} a_2^k(1-a_2)^{M_{\rm s}-k} P^{\rm p}({\rm Success}|k).
\end{equation}
A successful transmission for the PU occurs if the transmission rate used by the primary transmitter is less than or equal to the channel instantaneous capacity in a certain time slot, which occurs with probability $ P^{\rm p}({\rm Success}|k) $ when only $ k $ SUs are active. Next, we derive the SU success transmission probability, $ P^{\rm s}({\rm Success}|k) $, and the PU success transmission probability, $ P^{\rm p}({\rm Success}| k ) $. However, the complete derivation is given in the Appendix.

\begin{equation*}
P^{\rm s}({\rm Success}|k\!+\!{\rm PU})\!=\! \Pr\left \{ \tilde r_{\rm s}\!  \leq \!  \log_2(1+ \rm SINR) \!\right \}
\end{equation*}
\noindent
where SINR is the signal-to-interference-and-noise ratio. Due to symmetry, the probability of successful packet reception of the $j^{\it th}$ SU under interference of a set $\mathcal{K}$ of SUs depends on the number of SUs only. Assume that the active set is denoted by $\mathcal{A}\subseteq\{1,2,\dots,M_{\rm s}\}$ with cardinality $0\le \mathcal{K}\le M_{\rm s}$. Let $ r_{\rm s} \!=\! 2^{\tilde r_{\rm s}}-1 $. Hence, the probability of successful decoding of a packet at the $i^{th}$ SU's destination in case that the PU is active can be expressed as:

\begin{equation}
P^{\rm s}({\rm Success}|\mathcal{K}\!+\!{\rm PU})\!=\! \Pr\left \{\! r_{\rm s}\! <\! \frac{\gamma_2 h_{i,{\rm sd}_i}}{\mathcal{N}_\circ\!+\! \gamma_{\rm p}h_{{\rm p},{\rm sd}_i}\!+\!\sum_{\substack{{\theta\in\mathcal{A}}\\{\theta\ne i}}}\gamma_2h_{\theta,{\rm sd}_i}} \!\right \}
\label{hocko}
\end{equation}
\noindent
 For simplicity, we define $h_{\theta,{\rm sd}_i}=g_\theta$, $h_{i,{\rm sd}_i}=g_i$, and $h_{{\rm p},{\rm sd}_i}=g_{\rm p}$. The probability in (\ref{hocko}) can be rewritten as:
\begin{equation}
P^{\rm s}({\rm Success}|\mathcal{K}\!+\!{\rm PU})\!=\! \Pr\left \{\! r_{\rm s}\! <\! \frac{\gamma_2 g_i}{\mathcal{N}_\circ\!+\! \gamma_{\rm p}g_{\rm p}\!+\!\sum_{\substack{{\theta\in\mathcal{A}}\\{\theta\ne i}}}\gamma_2g_\theta} \!\right \}
\end{equation}
\noindent
If the PU is idle, we have
\begin{equation}
P^{\rm s}({\rm Success}|\mathcal{K})\!=\! \Pr\left \{\! r_{\rm s} \!<\! \frac{\gamma_1 g_i}{\mathcal{N}_\circ\!+\!\sum_{\substack{{\theta\in\mathcal{A}}\\{\theta\ne i}}}\gamma_1g_\theta} \!\right \}
\end{equation}
\noindent
Similarly, the correct packet reception of the PU is given by
\begin{equation}
P^{\rm p}({\rm Success}|\mathcal{K})\!=\! \Pr\left \{ r_{\rm p}\!<\! \frac{\gamma_{\rm p}g_p}{\mathcal{N}_\circ\!+\!\sum_{\substack{{\theta\in\mathcal{A}}}}\gamma_2g_\theta} \!\right \}
\end{equation}
\noindent
After the mathematical derivation given in the Appendix, the above probabilities are characterized as:
\begin{equation}\label{eqn_pr_s_k_PU}
P^{\rm s}({\rm Success}|\mathcal{K}+{\rm PU})\!=\! e^{-\!\delta_{\rm ss}\frac{r_{\rm s}\mathcal{N}_\circ}{\gamma_2}}\left ( \frac{1}{1+\frac{\delta_{\rm ss}r_{\rm s}\gamma_{\rm p}}{\gamma_2\delta_{\rm ps}}} \right ) \prod_{\substack{{\theta\in\mathcal{A}}}}\left (\! \frac{1}{1\!+\!r_{\rm s}} \!\right )
\end{equation}
\noindent
where $e^{-\!\delta_{\rm ss}\frac{r_{\rm s}\mathcal{N}_\circ}{\gamma_2}}$ represents the probability of successful secondary decoding when one SU is solely accessing the channel while PU is active. Since $1/(r_{\rm s}+1)$ is independent of $k$, the probability in (\ref{eqn_pr_s_k_PU}) is rewritten as

\begin{equation}
P^{\rm s}({\rm Success}|\mathcal{K}+{\rm PU})\!=\! e^{-\delta_{\rm ss}\frac{r_{\rm s}\mathcal{N}_\circ}{\gamma_2}}\left ( \!\frac{1}{1\!+\!\frac{\delta_{\rm ss}r_{\rm s}\gamma_{\rm p}}{\gamma_2\delta_{\rm ps}}} \!\right ) \left (\! \frac{1}{1\!+\!r_{\rm s}}\! \right )^{\mathcal{K}\!-\!1}
\end{equation}
\noindent
\begin{figure*}[!t]
\normalsize
\setcounter{equation}{14}
\begin{equation}\label{equation_long}
\begin{split}\mu_{\rm s} & =\sum _{\mathcal{K}=0}^{M_{\rm s}-1}{M_{\rm s} \choose \mathcal{K}}a_1^\mathcal{K}(1-a_1)^{(M_{\rm s}-\mathcal{K})}e^{-\delta_{\rm ss}\frac{r_{\rm s}\mathcal{N}_\circ}{\gamma_1}}\left (\frac{1}{1+r_{\rm s}}  \right )^{\mathcal{K}-1}
-\frac{\lambda_{\rm p} \sum _{\mathcal{K}=0}^{M_{\rm s}-1}{M_{\rm s} \choose \mathcal{K}}a_1^\mathcal{K}(1-a_1)^{(M_{\rm s}-\mathcal{K})}e^{-\delta_{\rm ss}\frac{r_{\rm s}\mathcal{N}_\circ}{\gamma_1}}\left (\frac{1}{1+r_{\rm s}}  \right )^{\mathcal{K}-1}}{\sum _{\mathcal{K}=0}^{M_{\rm s}}{M_{\rm s} \choose \mathcal{K}}a_2^\mathcal{K}(1-a_2)^{(M_{\rm s}-\mathcal{K})}e^{-\delta_{\rm pp}\frac{r_{\rm p}\mathcal{N}_\circ}{\gamma_{\rm p}}}\left (\frac{1}{1+\frac{\delta_{\rm pp}r_{\rm p}\gamma_2}{\delta_{\rm sp}\gamma_{\rm p}}}  \right )^{\mathcal{K}}} \\&
+ \frac{\lambda_{\rm p} \sum _{\mathcal{K}=0}^{M_{\rm s}-1}{M_{\rm s} \choose \mathcal{K}}a_2^\mathcal{K}(1-a_2)^{(M_{\rm s}-\mathcal{K})}e^{-\delta_{\rm ss}\frac{r_{\rm s}\mathcal{N}_\circ}{\gamma_2}}\left ( \frac{1}{1+\frac{\delta_{\rm ss}r_{\rm s}\gamma_{\rm p}}{\delta_{\rm ps}\gamma_2}} \right )\left (\frac{1}{1+r_{\rm s}}  \right )^{\mathcal{K}-1}}{\sum _{\mathcal{K}=0}^{M_{\rm s}}{M_{\rm s} \choose \mathcal{K}}a_2^\mathcal{K}(1-a_2)^{(M_{\rm s}-\mathcal{K})}e^{-\delta_{\rm pp}\frac{r_{\rm p}\mathcal{N}_\circ}{\gamma_{\rm p}}}\left (\frac{1}{1+\frac{\delta_{\rm pp}r_{\rm p}\gamma_2}{\delta_{\rm sp}\gamma_{\rm p}}}  \right )^{\mathcal{K}}} \end{split}
\end{equation}
\hrulefill
\vspace*{1pt}
\end{figure*}

Similarly,

\begin{equation}\label{eqn_pr_s_k}
\setcounter{equation}{11}
P^{\rm s}({\rm Success}|\mathcal{K})= e^{-\delta_{\rm ss}\frac{r_{\rm s}\mathcal{N}_\circ}{\gamma_1}} \prod _{\substack{{k=1}\\{k\ne i}}}^{\mathcal{K}} \left ( \frac{1}{1+r_{\rm s}} \right )
\end{equation}
Hence,
\begin{equation}
P^{\rm s}({\rm Success}|\mathcal{K})\!=\! e^{-\delta_{\rm ss}\frac{r_{\rm s}\mathcal{N}_\circ}{\gamma_1}} \left ( \!\frac{1}{1\!+\!r_{\rm s}} \!\right )^{\mathcal{K}\!-\!1}
\end{equation}
\noindent
In a similar fashion, the successful probability of a primary transmission when $\mathcal{K}$ SUs are active is given by
\begin{equation}\label{eqn_pr_p_k}
P^{\rm p}({\rm Success}|\mathcal{K})= e^{-\delta_{\rm pp}\frac{r_{\rm p}\mathcal{N}_\circ}{\gamma_{\rm p}}} \prod _{k=1}^{\mathcal{K}}\left ( \frac{1}{1+\frac{\delta_{\rm pp}r_{\rm p}\gamma_2}{\delta_{\rm sp} \gamma_{\rm p}}} \right )
\end{equation}
\noindent
Hence,
\begin{equation}
P^{\rm p}({\rm Success}|\mathcal{K})= e^{-\delta_{\rm pp}\frac{r_{\rm p}\mathcal{N}_\circ}{\gamma_{\rm p}}} \left ( \frac{1}{1+\frac{\delta_{\rm pp}r_{\rm p}\gamma_2}{\delta_{\rm sp} \gamma_{\rm p}}} \right )^{\mathcal{K}}
\end{equation}
\noindent
where $e^{-\delta_{\rm pp}\frac{r_{\rm p}\mathcal{N}_\circ}{\gamma_{\rm p}}}$ represents the probability of primary packet decoding when the PU is solely accessing the channel. Now, we can write the expression of the throughput of the $ i ^{th}$ SU by substituting from equation (\ref{eqn_mu_p}) into (\ref{eqn_pr_Q}) then, substitute from equation (\ref{eqn_pr_Q}), (\ref{eqn_pr_s_k_PU}), (\ref{eqn_pr_s_k}), and (\ref{eqn_pr_p_k}) into (\ref{eqn_mu_s}) to get (\ref{equation_long}) at the top of this page.


\section{Optimization Problem}\label{Optimization}

In this section, we present the optimization problem adopted in this paper. The SUs aim at maximizing their throughput given in (\ref{equation_long}) under stability constraint on the primary queue and average energy constraints on both the PU and the SU. The optimization problem is stated as follows.

\begin{equation}
\begin{aligned}
\setcounter{equation}{16}
& \underset{a_1,a_2,\gamma_1,\gamma_2}{\text{maximize}}
& & \mathrm{}\mu_{\rm s} \\
& \text{subject to}
& & \lambda_{\rm p}\leq \mu_{\rm p}\\
&&& E_{\rm p} \leq E_{\rm th,p}.
\\
&&& E_{\rm s} \leq E_{ \rm th,s}.
\label{opt}
\end{aligned}
\end{equation}
\noindent where $ E_{\rm s} $ and $E_{\rm p}$ are the secondary and the primary average transmit energy, respectively, $E_{\rm th,s}$ and $E_{\rm th,p}$ are the maximum allowable average transmit energy for the secondary and the primary terminals, respectively. The values of $E_{\rm th,s}$ and $E_{\rm th,p}$ depend on the application. The average transmit energy of the SU is given by

\begin{equation}
\begin{split}
E_{\rm s}= \Big[a_1 \gamma_1 (1-\frac{\lambda_{\rm p}}{\mu_{\rm p}}) +a_2\gamma_2 \frac{\lambda_{\rm p}}{\mu_{\rm p}}\Big] (T-\tau)
\end{split}
\end{equation}
The expression is explained as follows. If the PU is inactive, which occurs with probability $(1-\frac{\lambda_{\rm p}}{\mu_{\rm p}})$, each SU transmits with $ \gamma_1 (T-\tau)$ energy units with probability $a_1$, while if the PU is active, each SU transmits with $\gamma_2 (T-\tau)$ energy units with probability $a_2$.

Similarly, the average transmit energy of the PU is given by
\begin{equation}
\begin{split}
E_{\rm p}=  \gamma_{\rm p} T \frac{\lambda_{\rm p}}{\mu_{\rm p}}
\end{split}
\end{equation}


It should be pointed out here that the energy constraint on the PU is important to manage the average interference caused by the SUs on the PU. Specifically, if the average probability of successful decoding of the primary packets is decreased due to concurrent transmissions with the SUs, the probability of the primary queue to be non-empty will increase which, in turn, increases the average transmit primary energy. Based on that, having such constraint will always guarantee managing the access probabilities of the SUs such that the average transmit energy of the PU remains bounded.

The optimization problem (\ref{opt}) is solved numerically using MatLab \cite{wimob}. Since the optimization problem is generally nonconvex due to nonconvexity of the secondary mean service rate, $ \mu_{\rm s} $; the solver produces a local optimum solution. To enhance the quality and reliability of the solution and
increase the likelihood of obtaining the global optimum, the optimization problem is solved
many times, e.g., 1000 times, with different initializations of the decision variables. We note that since the access probabilities and power levels are obtained for a given average parameters of channels and arrival rates, as far as the average parameters are not changed, the difficulty of obtaining the optimization parameters are not high. More specifically, once we solve the problem for given parameters, the system can work for a long time using the obtained optimal parameters. More investigation of the optimization problem is part of our ongoing research. When the optimization problem is solved, the optimal access probabilities and the power levels are then announced to all SUs.

\begin{table*}
\centering
    \begin{tabular}{|c|c|c|c|c|}
 \hline
        $T=10^{-3} $ sec  &  $\tau=0.1T$  sec  & $b=10000 $ bits    & $W=10$ MHz     & $\mathcal{N}_\circ=10^{-11}$ Watts/Hz  \\ \hline
        $\delta_{\rm ss}=2$ &  $\delta_{\rm pp}=1$&  $\delta_{\rm ps}=2$ & $\delta_{\rm sp}=3$ & ~ \\ \hline
        $E_{\rm th,s} = 5 \times 10^{-5}$ joules& $E_{\rm th,p}=10^{-3}$ joules  & $ \gamma_1=2\times 10^{-10} $ Watts/Hz
        &  $ \gamma_2=1\times 10^{-10} $ Watts/Hz & \\ \hline
    \end{tabular}
\caption{Parameters' numerical values.}
 \label{Table1}
\end{table*}

\section{Numerical Results}\label{Results}

In this section, we present the results of our proposed scheme. We highlight our significant enhancement due to increasing the degrees of freedom of the optimized system. Table~I represents the numerical values of the system's parameters.

\begin{figure}
\centering
\includegraphics[width=0.9\columnwidth]{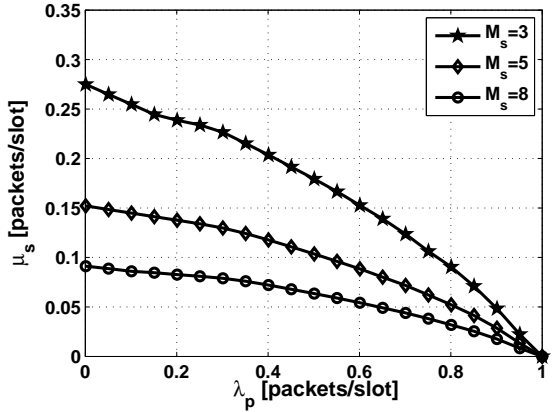}
\caption{Maximum secondary throughput for Different $ M_{\rm s}$.}\label{Stable_throughput_region}
\end{figure}

In order to show the gain of our proposed scheme, in Fig. \ref{Stable_throughput_region}, we plot the maximum secondary throughput for different numbers of the SUs. It is shown that as the total number of SUs, $ M_{\rm s} $, increases, the maximum secondary throughput for each SU decreases. This can also be conducted from equation (\ref{equation_long}), since the fraction of the access probabilities $ a_1 $ and $ a_2 $ are inversely proportional to $ M_{\rm s} $. Moreover, the transmission power levels, $ \gamma_1 $ and $ \gamma_2 $, assigned to SUs accessing the medium in a certain time slot are decreased in order to maintain an acceptable interference to the PU. For comparison purposes, we plot the case of conventional access scheme, where users access the channel with some probability when the PU is detected to be inactive. The power used by SUs in case of conventional access scheme is provided in Table \ref{Table1}. Note that the SUs remain idle when the PU is sensed to be active; hence, the transmit powers and the access probabilities of SUs are zero. The beneficial gain of the proposed protocol in case of adaptive and fixed powers is notable.

\begin{figure}
\centering
\includegraphics[width=0.9\columnwidth]{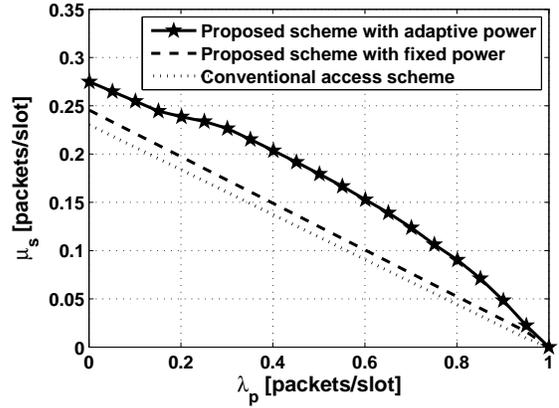}
\caption{Maximum secondary throughput for fixed and adaptive powers}\label{DOF}
\end{figure}

In Fig. \ref{DOF}, we illustrate the benefit we gain by increasing the system degrees of freedom by adding two additional optimization variables, $ \gamma_1 $ and $ \gamma_2$, rather than using a fixed power scheme (fixed power levels values are shown in table \ref{Table1}). The figure is generated with the parameters in Table \ref{Table1} and $M_{\rm s}=3$. The figure shows that, when we optimize over the transmission power levels we can increase the system throughput for certain PU load. The problem with the fixed power scheme is that, a SU may waste its opportunity to increase its transmission power when the PU can tolerate more interference due to existence of a small number of active SUs. On the other hand, if the fixed power levels were high and greedy to gain as much throughput as possible, SUs in this case will suffer higher outage probability causing a great loss to the transmitted packets, and hence, the advantage of our proposed scheme clearly appears.

\begin{figure}
\centering
\includegraphics[width=0.9\columnwidth]{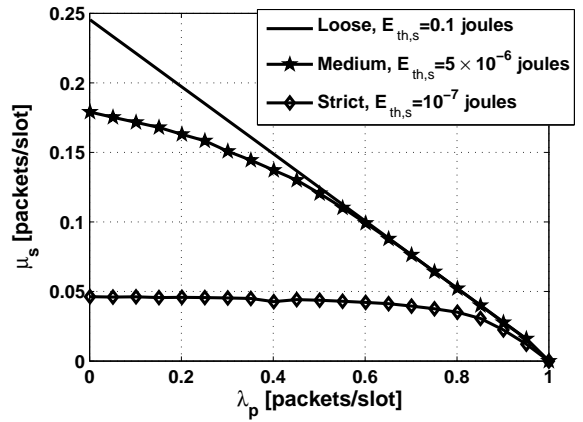}
\caption{Maximum secondary throughput for different SU energy constraints.}\label{Energy}
\end{figure}

Fig. \ref{Energy}, shows the effect of the SU average energy constraint, $ E_{\rm th,s} $, for three cases $ E_{\rm th,s}=0.1 $ joules, $ E_{\rm th,s}= 5 \times 10^{-6} $ joules and $ E_{\rm th,s}= 10^{-7} $ joules, to represent the loose, medium and strict constraints, respectively. It is clear that as the constraint becomes more strict, it reduces the maximum secondary throughput significantly. Adding an average energy constraint on the SU reduces the feasible set, of the four variables $ a_1 $, $ a_2 $, $ \gamma_1 $ and $ \gamma_2 $, over which the optimization is done. Therefore, decreasing $ E_{\rm th,s} $ (i.e, making the energy constraint more strict) causes a significant reduction in terms of the SU throughput.

\section{Conclusion}\label{Concl}

In this paper, we have investigated a cognitive network scenario with multiple SUs trying to randomly access one frequency channel. One of the main issues behind this work is addressing the energy efficiency of a cognitive setting. We have maximized the secondary throughput by using an adaptive power scheme while maintaining the average energy per user under certain threshold. As a fundamental point in the cognitive networks, guaranteeing the PU stability is also considered in the proposed scheme.

One possible extension of this work is to consider that the secondary transmitters aid the PU to deliver its packets through relaying. The possibility of having space-time coding at the SUs while relaying the primary packet can be also utilized.

\section*{Appendix}

Here, we give the details of the derivations of the probability of success, $ P^{\rm s}({\rm Success}|k) $, $ \Pr^{ \rm s}({\rm Success}|k+{\rm PU}) $ and $ \Pr^{ \rm p}({\rm Success}|k) $, used in equations (\ref{eqn_mu_s}) and (\ref{eqn_mu_p}). Assume that the active set is denoted by $\mathcal{A}\subseteq\{1,2,\dots,M_{\rm s}\}$ with cardinality $0\le \mathcal{K}\le M_{\rm s}$. The $i^{th}$ SU successful transmission probability when the PU is active is given by

\begin{equation}
P^{\rm s}({\rm Success}|\mathcal{K}\!+\!{\rm PU})\!=\! \Pr\left \{\! r_{\rm s}\! <\! \frac{\gamma_2 g_i}{\mathcal{N}_\circ\!+\! \gamma_{\rm p}g_{\rm p}\!+\!\sum_{\substack{{\theta\in\mathcal{A}}\\{\theta\ne i}}}\gamma_2g_\theta} \!\right \}
\end{equation}

\begin{equation}
P^{\rm s}({\rm Success}|\mathcal{K}\!+\!{\rm PU})\!=\! \Pr\left \{\! r_{\rm s}\! <\! \frac{\gamma_2 g_i}{\mathcal{N}_\circ\!+\! \gamma_{\rm p}g_{\rm p}\!+\!\sum_{\substack{{\theta\in\mathcal{A}}\\{\theta\ne i}}}\gamma_2g_\theta} \!\right \}
\end{equation}
\begin{equation}
=\Pr\left \{  g_i > \frac{r_{\rm s}}{\gamma_2}\left ( \mathcal{N}_\circ\!+\! \gamma_{\rm p}g_{\rm p}+\gamma_2\sum_{\substack{{\theta\in\mathcal{A}}\\{\theta\ne i}}}g_\theta \right ) \right \}
\end{equation}

\begin{equation}
\begin{split}
\!=\! & \int_{x_o\!=\!0}^\infty \int_{x_1\!=\!0}^\infty  \!\cdots\! \int_{x_{M_{\rm s}}\!=\!0}^\infty \int_{x_p\!=\!0}^\infty
e^{-\delta_{\rm ss}\left ( \frac{r_{\rm s}\mathcal{N}_\circ}{\gamma_2}+\!\frac{r_{\rm s}\gamma_{\rm p} x_p}{\gamma_2}\!+\!r_{\rm s}(x_0\!+\!\dots\!+\!x_{\mathcal{K}-1}) \right )} \\ &\prod_{k=1}^{\mathcal{K}-1}\left ( \delta_{\rm ss}e^{-\delta_{\rm ss}x_k} \right )\left ( \delta_{\rm ps}e^{-\delta_{\rm ps}x_p} \right ) dx_0dx_1...dx_{\mathcal{K}}dx_{\rm p}
\end{split}
\end{equation}
where $e^{(\cdot)}$ is the exponential function. Since the channel gains are i.i.d exponentially distributed random variables, which is the case when channels are Rayleigh fading, the above integration can be easily evaluated. After some mathematical manipulations, we get

\begin{equation}
P^{\rm s}({\rm Success}|\mathcal{K}+{\rm PU})\!=\! e^{-\!\delta_{\rm ss}\frac{r_{\rm s}\mathcal{N}_\circ}{\gamma_2}}\left ( \frac{1}{1+\frac{\delta_{\rm ss}r_{\rm s}\gamma_{\rm p}}{\gamma_2\delta_{\rm ps}}} \right ) \prod _{\substack{{k=1}\\{k\ne i}}}^{\mathcal{K}-1}\left (\! \frac{1}{1\!+\!r_{\rm s}} \!\right )
\end{equation}

In a similar fashion, we can characterize the other successful transmission probabilities.
\bibliographystyle{IEEEtran}
\bibliography{IEEEabrv,hossambib}

\begin{thebibliography}{10}
\providecommand{\url}[1]{#1}
\csname url@samestyle\endcsname
\providecommand{\newblock}{\relax}
\providecommand{\bibinfo}[2]{#2}
\providecommand{\BIBentrySTDinterwordspacing}{\spaceskip=0pt\relax}
\providecommand{\BIBentryALTinterwordstretchfactor}{4}
\providecommand{\BIBentryALTinterwordspacing}{\spaceskip=\fontdimen2\font plus
\BIBentryALTinterwordstretchfactor\fontdimen3\font minus
  \fontdimen4\font\relax}
\providecommand{\BIBforeignlanguage}[2]{{%
\expandafter\ifx\csname l@#1\endcsname\relax
\typeout{** WARNING: IEEEtran.bst: No hyphenation pattern has been}%
\typeout{** loaded for the language `#1'. Using the pattern for}%
\typeout{** the default language instead.}%
\else
\language=\csname l@#1\endcsname
\fi
#2}}
\providecommand{\BIBdecl}{\relax}
\BIBdecl

\bibitem{mitola1999cognitive}
J.~Mitola~III and G.~Maguire~Jr, ``Cognitive radio: making software radios more
  personal,'' \emph{Personal Communications, IEEE}, vol.~6, no.~4, pp. 13--18,
  1999.

\bibitem{haykin2005cognitive}
S.~Haykin \emph{et~al.}, ``Cognitive radio: brain-empowered wireless
  communications,'' \emph{IEEE journal on selected areas in communications},
  vol.~23, no.~2, pp. 201--220, 2005.

\bibitem{akyildiz2006next}
I.~Akyildiz, W.~Lee, M.~Vuran, and S.~Mohanty, ``Next generation/dynamic
  spectrum access/cognitive radio wireless networks: a survey,'' \emph{Computer
  Networks}, vol.~50, no.~13, pp. 2127--2159, 2006.

\bibitem{goldsmith2009breaking}
A.~Goldsmith, S.~A. Jafar, I.~Maric, and S.~Srinivasa, ``Breaking spectrum
  gridlock with cognitive radios: An information theoretic perspective,''
  \emph{Proceedings of the IEEE}, vol.~97, no.~5, pp. 894--914, 2009.

\bibitem{luo1999stability}
W.~Luo and A.~Ephremides, ``Stability of {N} interacting queues in
  random-access systems,'' \emph{IEEE Transactions on Information Theory},
  vol.~45, no.~5, pp. 1579--1587, Jul. 1999.

\bibitem{rao1988stability}
R.~Rao and A.~Ephremides, ``On the stability of interacting queues in a
  multiple-access system,'' \emph{IEEE Transactions on Information Theory},
  vol.~34, no.~5, pp. 918--930, Sep. 1988.

\bibitem{luo2006throughput}
J.~Luo and A.~Ephremides, ``On the throughput, capacity, and stability regions
  of random multiple access,'' \emph{IEEE Transactions on Information Theory},
  vol.~52, no.~6, pp. 2593--2607, 2006.

\bibitem{sadek}
A.~Sadek, K.~Liu, and A.~Ephremides, ``Cognitive multiple access via
  cooperation: protocol design and performance analysis,'' \emph{IEEE
  Transactions on Information Theory}, vol.~53, no.~10, pp. 3677--3696, Oct.
  2007.

\bibitem{el2011opportunistic}
A.~El-Sherif, A.~Sadek, and K.~Liu, ``Opportunistic multiple access for
  cognitive radio networks,'' \emph{IEEE Journal on Selected Areas in
  Communications}, vol.~29, no.~4, pp. 704--715, 2011.

\bibitem{TVT}
A.~El~Shafie and A.~Sultan-Salem, ``Stability analysis of an ordered cognitive
  multiple-access protocol,'' \emph{IEEE Transactions on Vehicular Technology},
  vol.~62, no.~6, pp. 2678--2689, 2013.

\bibitem{ghez1988stability}
S.~Ghez, S.~Verdu, and S.~Schwartz, ``Stability properties of slotted aloha
  with multipacket reception capability,'' \emph{IEEE Transactions on Automatic
  Control}, vol.~33, no.~7, pp. 640--649, 1988.

\bibitem{naware2005stability}
V.~Naware, G.~Mergen, and L.~Tong, ``Stability and delay of finite-user slotted
  {ALOHA} with multipacket reception,'' \emph{IEEE Transactions on Information
  Theory}, vol.~51, no.~7, pp. 2636--2656, July 2005.

\bibitem{erph}
S.~Kompella, G.~Nguyen, J.~Wieselthier, and A.~Ephremides, ``Stable throughput
  tradeoffs in cognitive shared channels with cooperative relaying,'' in
  \emph{Proceedings IEEE INFOCOM}, Apr. 2011, pp. 1961--1969.

\bibitem{wimob}
A.~El~Shafie and A.~Sultan, ``Optimal random access and random spectrum sensing
  for an energy harvesting cognitive radio,'' in \emph{IEEE 8th International
  Conference on Wireless and Mobile Computing, Networking and Communications
  (WiMob)}, 2012, pp. 403--410.

\bibitem{ourletter}
A.~{El Shafie} and A.~Sultan, ``Optimal random access for a cognitive radio
  terminal with energy harvesting capability,'' \emph{IEEE Communications
  Letters}, vol.~17, no.~6, pp. 1128--1131, 2013.

\bibitem{grace2009using}
D.~Grace, J.~Chen, T.~Jiang, and P.~D. Mitchell, ``Using cognitive radio to
  deliver �green�communications,'' in \emph{IEEE International Conference
  on Cognitive Radio Oriented Wireless Networks and Communications}, 2009, pp.
  1--6.

\bibitem{feng2012survey}
D.~Feng, C.~Jiang, G.~Lim, L.~Cimini~Jr, G.~Feng, and G.~Li, ``A survey of
  energy-efficient wireless communications,'' \emph{IEEE Communications Surveys
  Tutorials}, 2012.

\bibitem{lim2012energy}
G.~Lim and L.~Cimini~Jr, ``Energy-efficient cooperative relaying in
  heterogeneous radio access networks,'' \emph{IEEE Wireless Communications
  Letters}, 2012.

\bibitem{hossain2012green}
E.~Hossain, V.~K. Bhargava, and G.~P. Fettweis, \emph{Green radio communication
  networks}.\hskip 1em plus 0.5em minus 0.4em\relax Cambridge University Press,
  2012.

\bibitem{grant2008cvx}
M.~Grant, S.~Boyd, and Y.~Ye, ``Cvx: Matlab software for disciplined convex
  programming,'' 2008.

\end{thebibliography}
\balance

\end{document}